\documentclass[a4paper,11pt]{article}
\pdfoutput=1

\usepackage[a4paper,left=2.73cm,right=2.7cm,top=3cm,bottom=3.5cm]{geometry}
\usepackage[mathscr]{eucal}
\usepackage{color,amsmath,amssymb,amsfonts}
\usepackage{enumerate}
\usepackage{hyperref}
\usepackage{graphicx}
\usepackage{psfrag}
\usepackage{dcolumn,bm}
\usepackage{mdframed}

\usepackage[utf8]{inputenc}

\newcommand{\bea}{\begin{eqnarray}}
\newcommand{\beal}[1]{\begin{eqnarray}\label{#1}}
\newcommand{\eea}{\end{eqnarray}}
\newcommand{\be}{\begin{equation}}
\newcommand{\bel}[1]{\begin{equation}\label{#1}}
\newcommand{\ee}{\end{equation}}

\newcommand{\bit}{\begin{itemize}}
\newcommand{\eit}{\end{itemize}}
\newcommand{\ben}{\begin{enumerate}}
\newcommand{\een}{\end{enumerate}}

\def\d{\partial}

\newcommand{\mt}[1]{\textrm{\tiny #1}}
\newcommand{\nc}{N_\mt{c}}
\newcommand{\nf}{N_\mt{f}}
\newcommand{\fsub}{\mt{f}}
\newcommand{\bsub}{\mt{b}}

\newcommand{\sac}{\, , \qquad}

\newcommand{\eqq}[1]{(\ref{#1})}
\newcommand{\fig}[1]{Fig.~\ref{#1}}

\newcommand{\bal}{\begin{align}}
\newcommand{\eal}{\end{align}}
\newcommand{\bse}{\begin{subequations}}
\newcommand{\ese}{\end{subequations}}

\newcommand{\Mq}{M_\mt{q}}
\newcommand{\mui}{\mu_\mt{I}}

\newcommand{\Nu}{n_\mt{u}}
\newcommand{\Nd}{n_\mt{d}}
\newcommand{\Nr}{n_\mt{R}}
\newcommand{\Ni}{n_\mt{I}}

\newcommand{\fmag}{F_\mt{mag}} 

\def\d{\mathrm{d}}
\newcommand{\tr}{\mbox{Tr}}

\begin{document}

\begin{titlepage}

\thispagestyle{empty}

\begin{flushright}
\hfill{ICCUB-18-016}
\end{flushright}

\vspace{40pt}  
	 
\begin{center}

{\LARGE \textbf{A Supersymmetric Color Superconductor from Holography}}
	\vspace{30pt}
		
{\large \bf Ant\'on F. Faedo,$^{1}$   David Mateos,$^{1,\,2}$   \\ [2mm]
Christiana Pantelidou$^{3}$  and Javier Tarr\'\i o$^{4}$}

\vspace{25pt}

{\normalsize  $^{1}$ Departament de F\'\i sica Qu\'antica i Astrof\'\i sica and Institut de Ci\`encies del Cosmos (ICC),\\  Universitat de Barcelona, Mart\'\i\  i Franqu\`es 1, ES-08028, Barcelona, Spain.}\\
\vspace{15pt}
{ $^{2}$Instituci\'o Catalana de Recerca i Estudis Avan\c cats (ICREA), \\ Passeig Llu\'\i s Companys 23, ES-08010, Barcelona, Spain.}\\
\vspace{15pt}
{ $^{3}$Centre for Particle Theory and Department of Mathematical Sciences, Durham University, Durham, DH1 3LE, U.K.}\\
\vspace{15pt}
{ $^{4}$
Physique Th\'eorique et Math\'ematique, Universit\'e Libre de Bruxelles (ULB) \\
and International Solvay Institutes, Campus de la Plaine CP 231, B-1050, Brussels, Belgium.}

\vspace{40pt}
				
\abstract{
We use holography to study $d=4$, $\mathcal{N}=4$, SU($\nc$) super Yang-Mills coupled to \mbox{$\nf \ll \nc$} quark flavors. We place the theory at finite isospin  density $\Ni$  by turning on an isospin chemical potential  $\mui=\Mq$, with $\Mq$ the quark mass. We also turn on two \mbox{R-symmetry} charge densities $n_1=n_2$. We show that the ground state is a supersymmetric, superfluid, color superconductor, namely a finite-density state that preserves a fraction of supersymmetry in which part of the global  symmetries and part of the gauge symmetries are spontaneously broken. The holographic description consists of $\nf$ D7-brane probes in $\mbox{AdS}_5 \times \mbox{S}^5$.  The symmetry  breaking is due to the dissolution of some D3-branes inside the D7-branes triggered by the electric field associated to the isospin charge. The massless spectrum contains Goldstone bosons and their fermionic superpartners. The massive spectrum contains long-lived, mesonic quasi-particles if $\Ni \ll \mui^3$, and no quasi-particles otherwise. We discuss the possibility that, despite the presence of mass scales and charge densities in the theory, conformal and relativistic   invariance arise as emergent symmetries in the infrared. 
}

\end{center}

\end{titlepage}

\tableofcontents

\hrulefill
\vspace{10pt}

\section{Introduction}
Quantum Chromodynamics (QCD) at non-zero baryon density $n_\bsub$ is notoriously difficult to analyze. Because of asymptotic freedom, the preferred phase 
at asymptotically high density can be shown to be a color-flavor locked (CFL) configuration \cite{Alford:1997zt,Alford:1998mk} (for a review see \cite{Alford:2007xm}). The ground state in this regime is a color superconductor, namely a finite-density state in which the color symmetry is Higgsed. Following a common abuse of language, we will refer to this as the spontaneous breaking of the color symmetry. In addition, the CFL ground state is also a superfluid, since the baryon number symmetry is spontaneously broken too. In the regime of high but finite density, such as at the core of neutron stars, no first-principle  calculations are possible. The only non-perturbative tool, namely lattice QCD, is of limited applicability due to the so-called sign problem \cite{deForcrand:2010ys}. 

This situation provides one motivation to study the physics of QCD as some other conserved charge is taken to be large, for example the isospin charge. In this case the sign problem is absent and the theory can be simulated on the lattice (see e.g.~\cite{Kogut:2004zg}). Analytical methods can also be used \cite{Son:2000xc,Son:2000by}. The emergent picture is that the ground state is a superfluid with superfluidity driven by a pion condensate at low density and by a quark-antiquark condensate at high density. No color superconductivity was found in these analysis. 

In this paper we give a step towards the holographic description of color superconducting phases. In this context the goal is not to do precision physics but to perform first-principle calculations that may lead to interesting insights \cite{Mateos:2011bs}. In the case of QCD at non-zero temperature, the insights obtained through this program range from static properties to far-from-equilibrium dynamics of strongly coupled plasmas (see e.g.~\cite{CasalderreySolana:2011us} and references therein).

We will investigate a simple yet extremely rich holographic model which exhibits both color superconductivity and superfluidity when a certain combination of conserved charges is taken to be large. Our model differs from QCD in many respects, including the fact that it is supersymmetric, that it exhibits no chiral symmetry and hence no pions, and that it possesses an R-symmetry that is absent in QCD. Therefore we do not claim that our results have any direct implications for real-world QCD.  However, we believe that they are interesting for three reasons. First, they show that color superconductivity does appear in holography when some conserved charges are large (in this case a combination of isospin charge and R-charge). Second, we expect that a similar holographic mechanism will give rise to color superconductivity in the presence of baryon density \cite{progress}. Third, to the best of our knowledge  our model is the first example of a supersymmetric color superconductor. 
Color superconductivity in supersymmetric theories has been previously considered in e.g.~\cite{Harnik:2003ke,Arai:2005pk,Rajput:2011zzc}, but in these cases all the supersymmetries are broken by the ground state. In contrast, in our model the ground state leaves some supersymmetry unbroken. 
We expect that this property will facilitate  a precise comparison between the strong-coupling limit described by holography and the weak-coupling regime accessible via perturbative field theory methods.

Color superconductivity in the holographic context has been previously explored. Refs.~\cite{Chen:2009kx,Rozali:2012ry} considered baryon density instead of isospin density, Ref.~\cite{Basu:2011yg} studied a bottom-up model instead    of a top-down model, and 
Refs.~\cite{BitaghsirFadafan:2018iqr,Ghoroku:2019trx}  mimicked  the breaking of the color symmetry as the breaking of a global symmetry.

\section{Model}
Type IIB string theory on the near-horizon geometry of $\nc$ D3-branes and $\nf$ \mbox{D7-branes} is dual to $d=4$, $\mathcal{N}=4$, SU($\nc$) super Yang-Mills theory coupled to $\nf$ hypermultiplets in the fundamental representation. The presence of the hypermultiplets breaks supersymmetry  to $\mathcal{N}=2$, so we will refer to this theory simply as ``the  $\mathcal{N}=2$ gauge theory''. Although the hypermultiplets contain both bosons and fermions, we will loosely refer to them as ``flavors'' or ``quarks''. In the regime $\nf\ll \nc$ the D7-branes can be treated as probes \cite{Karch:2002sh}  in the $\mbox{AdS}_5 \times \mbox{S}^5$ geometry 
\be\label{eq.metric10d}
\d s^2 = H^{-\frac{1}{2}} \left( -\d t^2 + \d {\vec x}^2 \right) + 
H^{\frac{1}{2}} \left( \d y_i^2 + \d z_\alpha^2  \right), \,\,\,\,\,\,\,\,\,
\ee
where $t,\vec x$ 
are the four gauge theory directions parallel to the D3-branes, $y^i$ with $i=1, \ldots, 4$ are the coordinates along the D7-branes orthogonal to the D3-branes, and $z^\alpha$ with $\alpha=1,2$ are the coordinates orthogonal to both sets of branes. We will often write the metric in the  $y^i$ directions in spherical coordinates as
\be
\label{s3}
\d y_i^2 = dr^2 + r^2 \left(  \omega_1^2 + \omega_2^2 + \omega_3^2 \right) \,,
\ee
where $\omega_n$ are the left-invariant forms on S$^3$.
$H$ is the usual harmonic function in the six-dimensional space transverse to the D3-branes:
\be
\label{H}
H = \frac{L^4}{\left( r^2 + z_1^2 + z_2^2 \right)^2}\,,
\ee
with $L$ the radius of $\mbox{AdS}_5$ and  $\mbox{S}^5$. 

The dynamics of the $\nf$ D7-branes may be described by the non-Abelian action of \cite{Myers:1999ps}. At the lowest order in the string tension this reduces to a super-Yang-Mills-Higgs (SYMH) action 
together  with extra couplings to background fluxes coming from the Wess-Zumino term. As emphasised by the author himself, the action in \cite{Myers:1999ps} is known to be incomplete,  but it  seems to capture the exact physics for supersymmetric configurations \cite{Hashimoto:1997px,Bak:1998xp}. In fact, supersymmetric solutions of the SYMH action often become  solutions of the full action. This is also the case here \cite{appear}, and therefore we will effectively work with the SYMH action. For simplicity  we will focus on the case $\nf=2$ and we will refer to the two  flavors as $u$ and $d$ quarks. 

In the  background \eqq{eq.metric10d} the SYMH action takes the form
\bea\label{eq.SYMHaction}
\frac{S}{T_\mt{D7}} &=& -  \int   \frac{1}{2} \tr \Big(  F \wedge * F  + H^{\frac{1}{2}}  \delta_{\alpha\beta} DZ^\alpha \wedge * D Z^\beta  \Big)
 \nonumber \\ 
&& - \int\frac{1}{2}  H^{-1} \d t \wedge \d^3 x \wedge \tr \left( F \wedge F \right)
\,. \,\,\,\,\,\,
\eea
Throughout this paper we set $2\pi\ell_s^2=1$, so all quantities are effectively dimensionless.  
The last term in \eqq{eq.SYMHaction} comes from the coupling to the RR five-form that supports the geometry \eqq{eq.metric10d}. $F$ is the U(2)$_\fsub$ non-Abelian field strength on the world volume of the D7-branes and $D$ is the gauge covariant derivative. $Z^\alpha$ are non-Abelian scalars (Higgs fields) in the adjoint of U(2)$_\fsub$, which parametrize the (in general non-commuting) positions of the D7-branes in the $z^\alpha$-plane. The Hodge dual $*$ is taken with respect to the eight-dimensional induced metric on the branes, $g$. The solutions that we will consider will all lie within the SU(2)$_\fsub$ subgroup of 
\mbox{$\mbox{U(2)}_\fsub = \mbox{SU(2)}_\fsub \times \mbox{U(1)}_\bsub$}   and  will be translationally-invariant along the $\vec{x}$-directions. Therefore we will be effectively studying SU(2)$_\fsub$ configurations in the five  dimensions $\{t, \vec{y} \}$. The $\mbox{U(1)}_\mt{b}$ charge can be thought of as the baryon number and will play no role here.

\section{Higgs branch}
\label{higgs}
The $\mathcal{N}=2$ gauge theory possesses a continuous moduli space of vacua parametrized by the vacuum expectation values (VEVs) of  (s)quark bilinear operators. Supersymmetry guarantees that all the ground  states in this  so-called Higgs branch are exactly degenerate.\footnote{There is also a Coulomb branch parametrized by the VEVs of the  adjoint scalars of the theory.}   Holographic studies of the Higgs branch include \cite{Erdmenger:2005bj,Guralnik:2004ve,Guralnik:2004wq,Guralnik:2005jg}. In  these papers no baryon or isospin density was considered. 

At a generic vacuum on the Higgs branch part of the SU($\nc$) gauge symmetry is spontaneously broken (some global symmetries are broken too, see Sec.~\ref{symm}).  At weak coupling in the gauge theory the breaking  can be seen by analysing the part of the SU($\nc$) symmetry that is broken by a specific set of VEVs in a fixed gauge (see e.g.~\cite{Erdmenger:2005bj}). At strong coupling the breaking is due to the separation of some D3-branes from the others, which 
gives a mass to the strings stretching between them. Since these strings are dual to the  gluons in the gauge theory this signals the breaking of the 
SU($\nc$) gauge group. The separated D3-branes dissolve inside the D7-branes and appear as instantons of the SYMH theory on the D7-branes.\footnote{This is a particular example of the more general phenomenon of ``Branes within branes'' \cite{Witten:1995im,Douglas:1995bn} whereby a low-dimensional brane dissolves inside a high-dimensional brane. From the viewpoint of the effective field theory on the high-dimensional brane, the low-dimensional brane appears as a flux of the worldvolume gauge field. A pedagogical discussion can be found in \cite{Johnson:2000ch}. In the context of holography only the case of an instanton flux on the D7-branes leads to color superconductivity because this is the case where the dissolved objects are precisely the D3-branes associated to the color symmetry of the dual gauge theory.} 

The breaking can also be seen explicitly by considering the backreaction of the D7-branes-plus-instanton on the spacetime fields. This results in a position-dependent RR five-form flux \cite{Faedo:2016jbd} and therefore in a scale-dependent effective rank of the gauge group. The size of the instanton  
$\Lambda$ is dual to the scale of gauge symmetry breaking and, in the absence of charge densities, it is arbitrary.  A key point of our paper is that, in the presence of  isospin and R-charge densities, the scale of gauge symmetry breaking is fixed by the charges. In other words, in the absence of charge densities one may place the theory at an arbitrary point on the Higgs branch and thus break the gauge symmetry ``by hand''. In contrast, in the presence of charges the system is driven  to a specific point on the Higgs branch fixed by the charges and the gauge symmetry is broken dynamically.

\section{Solution}
We consider  a direct importation of the dyonic instanton solution of \cite{Lambert:1999ua}
\bse
\label{confi}
\bal
\label{last}
A &= a_t (r) \, \d t \otimes  \sigma^3 + a(r) \, \delta_{mn}\,  \omega^n \otimes \sigma^n \,, \\[2mm]
Z^1 & = Z =  \phi(r) \, \sigma^3 \sac Z^2 = 0 \,,
\label{differ}
\end{align}
\ese
where $\sigma^n$ are the Pauli matrices.
We split the field strength into purely electric and purely magnetic parts:  $F=\d t\wedge E+\fmag$.
If the following first-order BPS conditions are satisfied 
\be
\label{BPS}
F_\mt{mag} = - \star F_\mt{mag} \sac E = - DZ \,,
\ee
where the Hodge dual $\star$ is taken with respect to the flat metric along the $y$-directions,  then the configuration \eqq{confi} preserves ${\mathcal N}=1$  supersymmetry and it solves the second-order equations of motion of  SYMH theory in flat space \cite{Lambert:1999ua}. The first equation is the usual (anti)selfduality condition associated to instantonic configurations, whereas the second one relates the electric field and the scalar and, via Gauss' law, implies that $D\star DZ=0$. As expected from supersymmetry, 
this  configuration also solves the equations of motion of the full action for a pair of D-branes  \cite{Myers:1999ps}  in flat space \cite{Zamaklar:2000tc}.  Crucially, the same is true for a pair of D7-branes placed 
 in the background \eqq{eq.metric10d} \cite{appear}.

The solution of the BPS equations of interest to us is
\be
\label{solution}
a(r) = \frac{\Lambda^2}{r^2 + \Lambda^2} \sac
a_t (r)= \phi(r) = \frac{\Mq r^2}{r^2 + \Lambda^2} \,,
\ee
where $\Lambda$ and $\Mq$ are integration constants. The solution is completely regular everywhere. The asymptotic form 
\be
\label{asym}
a_t (r)= \phi(r) \simeq \Mq -  \frac{\Mq \Lambda^2}{r^2} + \cdots
\ee
will be useful below. 

Choosing the scalar to point in the  $\sigma^3$ direction explicitly breaks the SU(2)$_\fsub$  gauge symmetry down to U(1)$_\fsub$. This allows us to define the electric charge of the instanton by projecting onto the unbroken U(1)$_\fsub$ as
\be
q= \lim_{r\to\infty }\frac{1}{\Mq} \int_{S^3} r^3 \tr \left( Z  E_r  \right)  =
2\pi^2 \Lambda^2  \Mq  \,,
\ee
where the integral is taken on a three-sphere of radius $r$ in the metric \eqq{s3}. 
Despite the explicit breaking of SU(2)$_\fsub$ to an Abelian subgroup, the instanton is prevented from collapse by the  non-zero angular momentum produced by the crossed electric and magnetic non-Abelian fields \cite{Eyras:2000dg}. The Poynting momentum density is aligned with the $\omega^3$ left-invariant form in \eqq{s3}, which results in a self-dual angular momentum with equal skew-eigenvalues  in the $y^i$-directions $n_1=n_2\propto q$.

\section{Physical interpretation} \label{phys}

\begin{figure}[t]
\begin{center}
\includegraphics[width=.4\textwidth]{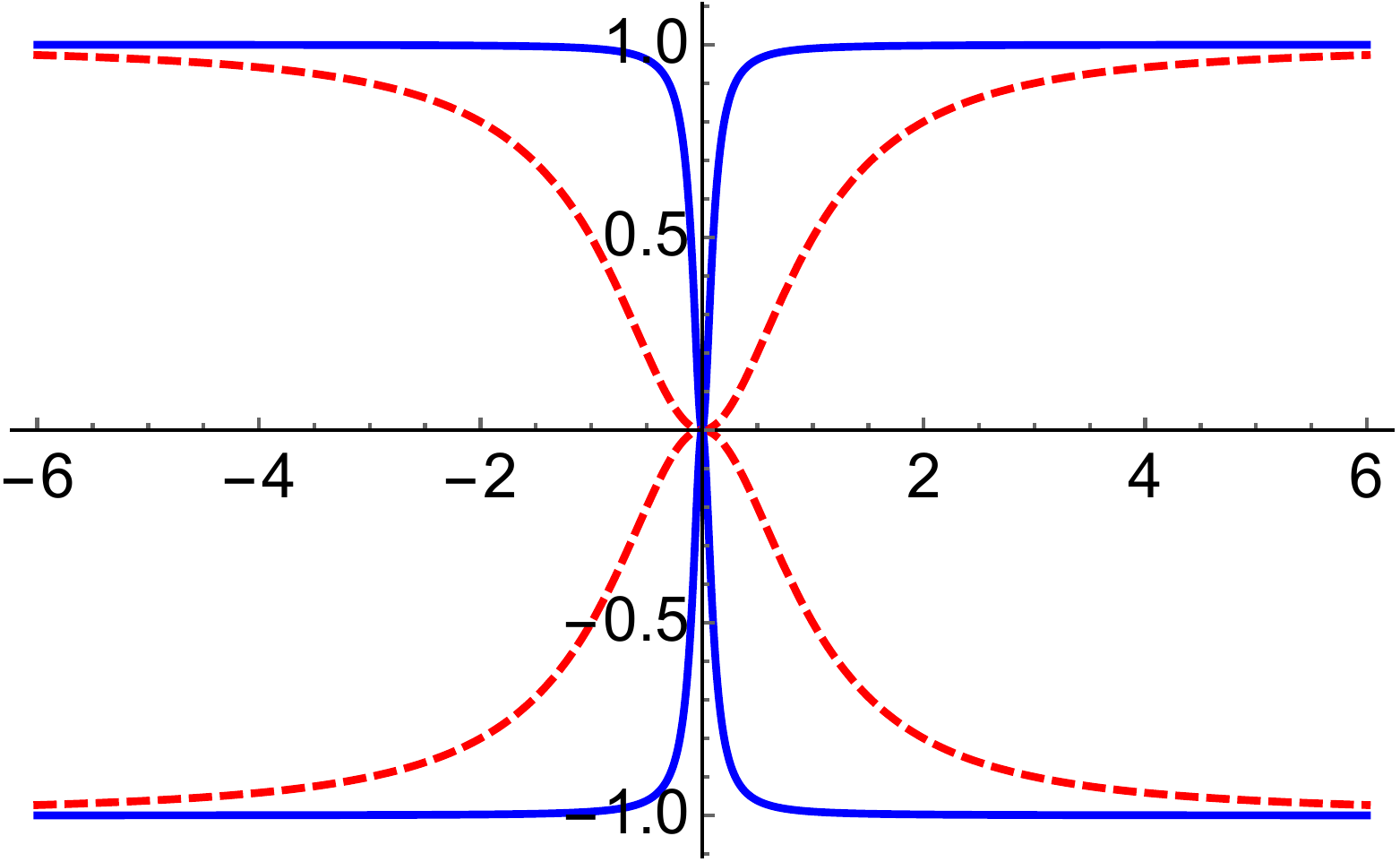} 
\put(-124,132){\mbox{\Large $\pm \phi/\Mq$}}
\put(5,60){\mbox{\Large $r$}}
\end{center}
\vspace{-5mm}
\caption{\small Brane profiles  for  
$\Lambda=1$ (dashed, red curve) and $\Lambda=1/10$ (solid, blue curve).
}
\label{embedding}
\end{figure}

The fact that $Z(r)$ is proportional to $\sigma^3$, which is diagonal with entries $\pm 1$,  means that the branes bend in opposite directions along the \mbox{$z^1$-axis} with otherwise identical profiles, as shown in \fig{embedding}.  
The asymptotic behaviour \eqq{asym}  has two immediate consequences. First,  
the constant $\Mq$ corresponds to the quark mass \cite{Karch:2002sh}.
To be precise, the quark mass is a complex number and we see that the masses of the $u$ and $d$ quarks are equal in magnitude but have opposite phases. The same is true for the corresponding quark condensates 
\be
\langle \bar u u \rangle = - \langle \bar d d \rangle \propto - \Mq \Lambda^2\,. 
\ee
Note that this corresponds to a vanishing chiral condensate
\be
\langle \bar \Psi \Psi \rangle = \langle \bar u u +  \bar d d \rangle = 0 
\ee
but to a nonzero condensate of the form 
\be
\langle \bar \Psi \sigma^3 \Psi \rangle = \langle \bar u u -  \bar d d \rangle 
\propto -  \Mq \Lambda^2 \,.
\ee
Second, 
the isospin chemical potential and the charge densities are given by \cite{Apreda:2005yz,Kobayashi:2006sb}
\be
\label{through1}
\mui= \Mq \sac \Ni = \Nu = -\Nd \propto - \Mq \Lambda^2  \,.
\ee
From the viewpoint of the dual gauge theory, the angular momentum  corresponds to equal R-charge densities \cite{Gubser:1998jb,Chamblin:1999tk,Cvetic:1999ne}
\be
\label{through2}
\Nr \equiv n_1=n_2
\ee
along two of the three U(1) factors in the Cartan subalgebra of the SO(6) R-symmetry of   $\mathcal{N}=4$ SYM. Note that all the charge densities  are comparable since 
\be
\label{compa}
\Ni \sim \Nr \sim q\,.
\ee
Moreover, the size of the instanton is not a free parameter but is fixed 
as
\be
\label{givenby}
\Lambda^2 \propto \frac{\Nr}{\mui} \,.
\ee
In summary, the solution \eqq{solution} represents a state in the 
$\mathcal{N}=2$ gauge theory with non-zero isospin and R-charge densities related to one another and to the quark mass through 
\eqq{through1}, \eqq{through2} and \eqq{compa}. The 
$\mathcal{N}=2$ gauge theory has been previously studied both at non-zero isospin density \cite{Apreda:2005yz,Erdmenger:2007ap,Erdmenger:2008yj} and at non-zero baryon density \cite{Kobayashi:2006sb,Karch:2007pd,Karch:2007br,Mateos:2007vc,Chen:2009kx}.  In  these papers no supersymmetric ground state was identified. 

\section{Symmetry breaking}\label{symm}

Four-dimensional \mbox{$\mathcal{N}=4$} SYM is invariant under conformal transformations and  possesses an SO(6) 
R-symmetry. Coupling the theory to massive quarks has several effects. First, it breaks conformal invariance. Second, it decreases  the supersymmetry to 
$\mathcal{N}=2$. Third,  it reduces the SO(6) symmetry to \mbox{$\mbox{SO(4)} = \mbox{SU(2)}_\mt{L} \times \mbox{SU(2)}_\mt{R}$}, where $\mbox{SU(2)}_\mt{R}$ is the \mbox{R-symmetry} of the 
$\mathcal{N}=2$ algebra and $\mbox{SU(2)}_\mt{L}$ is a global symmetry that does not act on the $\mathcal{N}=2$ supercharges. For massless quarks there would be an extra U(1)$_\mt{R}$ factor. The addition of $\nf=2$ quark flavors of equal mass would introduce a new 
$\mbox{SU(2)}_\fsub \times \mbox{U(1)}_\bsub$ global symmetry that would rotate the quarks into one another. If the quark masses differ even just in phase, as in our case, then this symmetry is explicitly broken to 
$\mbox{U(1)}_\fsub \times \mbox{U(1)}_\bsub$. 

On the gravity side these breakings can be understood geometrically. The original SO(6) symmetries are  the isometries of the S$^5$. The addition of the D7-branes selects a four-plane in the $\mathbb{R}^6$ space transverse to the D3-branes and thus reduces the symmetry to SO(4). The $\mbox{SU(2)}_\mt{L}$ factor leaves each $\omega_n$ form in \eqq{s3}  invariant, whereas the $\mbox{SU(2)}_\mt{R}$ rotates them into one another. For massless quarks the branes lie at the origin of the \mbox{$z^{12}$-plane} and an additional U(1)$_\mt{R}$ corresponding to rotations in this plane is preserved. Instead, a non-zero quark mass  breaks the  U(1)$_\mt{R}$ explicitly. The $\mbox{SU(2)}_\fsub \times \mbox{U(1)}_\bsub$ global symmetry of the gauge theory becomes the non-Abelian gauge symmetry on the pair of D7-branes, which is broken explicitly to 
$\mbox{U(1)}_\fsub \times \mbox{U(1)}_\bsub$ by the fact that the branes bend in opposite direction in the $z^1$-axis. 

In summary, \mbox{$\mathcal{N}=4$} SYM coupled to massive quarks of unequal masses is invariant under an SU($\nc$) gauge symmetry and an 
\mbox{$\mbox{SU(2)}_\mt{L} \times \mbox{SU(2)}_\mt{R} \times 
\mbox{U(1)}_\fsub \times \mbox{U(1)}_\bsub$} global symmetry. Placing the 
theory  at non-zero  density by turning on $\Ni$ and $\Nr$ results in the spontaneous breaking of some of these symmetries. This can be easily seen on the gravity side. First, the presence of an 
instanton indicates that some D3-branes have dissolved inside the D7-branes. As explained in Sec.~\ref{higgs} this implies that the gauge group is spontaneously broken, 
with the scale of the breaking  given by \eqq{givenby}. Second, the self-dual angular momentum generated by the simultaneous  presence of the isospin electric field and the instanton spontaneously breaks part of the global symmetries as ${\mbox{SU(2)}_\mt{R}\times\mbox{U(1)}_\fsub\to\mbox{U(1)}_\mt{D}}$, where the group on the right-hand side is the diagonal U(1) in the Cartan subalgebra of the left-hand side. This breaking can be seen geometrically as follows. The fact that the Poynting vector distinguishes between $\omega^3$ and $\omega^{1,2}$ breaks $\mbox{SU(2)}_\mt{R}$ to the $\mbox{U(1)}_\mt{R}$ subgroup that rotates $\omega^{1,2}$ into one another. The fact that only simultaneous rotations of $\omega^n$ and $\sigma^n$ leave the second term in \eqq{last} invariant then breaks $\mbox{U(1)}_\mt{R} \times  \mbox{U(1)}_\fsub$ to the diagonal $\mbox{U(1)}_\mt{D}$. This  is similar to the breaking  of $\mbox{SU(2)}_\mt{R} \times \mbox{SU(2)}_\fsub$ to $\mbox{SU(2)}_\mt{D}$ that takes  place on the Higgs branch  in the absence of the electric field (see e.g.~\cite{Erdmenger:2005bj}). 

\section{Spectrum}\label{spec}

Since several global symmetries are broken spontaneously, we expect the spectrum to contain massless bosons. 
Given that  the ground state preserves $\mathcal{N}=1$ supersymmetry, we expect them to be accompanied by the corresponding massless, fermionic  superpartners. 
The fact that the ground state 
breaks Lorentz invariance means that the number of these modes need not coincide with the number of broken generators \cite{Halperin,Nielsen:1975hm}. Holographic examples of this phenomenon include \cite{Filev:2009xp,Amado:2013xya,Argurio:2015via}. We will report on the massless modes elsewhere \cite{appear}.

\begin{figure}[t]
\begin{center}
\includegraphics[width=.4\textwidth]{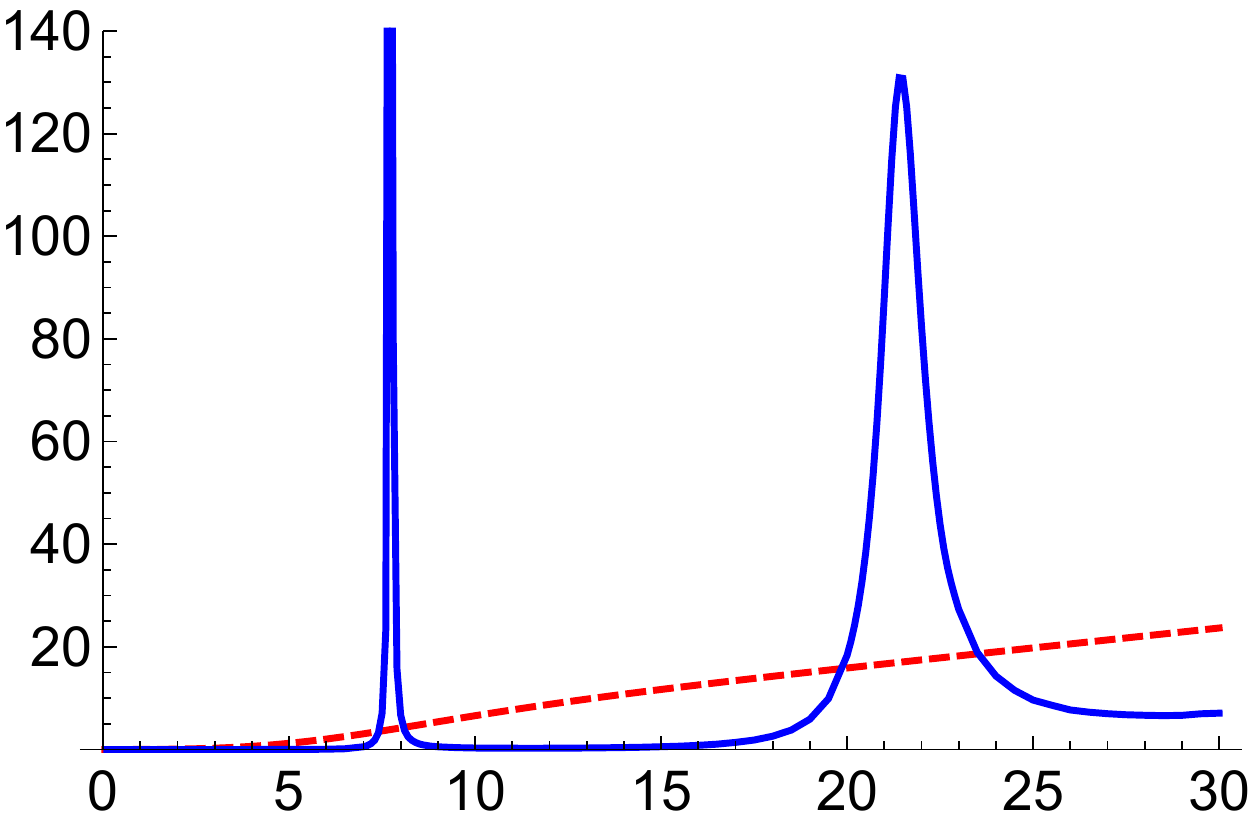} 
\put(-200,70){\mbox{\Large $\chi$}}
\put(-110,-15){\mbox{\large $L^4 \omega^2 / M_q^2$}}
\end{center}
\vspace{-5mm}
\caption{\small Spectral function described in the text (with arbitrary normalization)  for $\epsilon=1$ (dashed, red curve) and 
$\epsilon=1/100$ (solid, blue curve).
}
\label{spectrum}
\end{figure}

The qualitative properties of the spectrum of massive modes depend on the value of the ratio ${\epsilon=\Lambda^2/\Mq^2\propto\Ni/\mui^3}$. 
For small $\epsilon$ the spectrum contains long-lived, mesonic quasi-particles; otherwise no massive quasi-particles are present. This is illustrated in \fig{spectrum}, where we show the spectral function at zero spatial momentum for the gauge theory operator dual to the $\sigma^3$ component of the scalar field $Z^2$,  calculated via standard methods \cite{Son:2002sd} from fluctuations governed  by the action \eqq{eq.SYMHaction}. For small $\epsilon$ we see high and narrow peaks associated to long-lived excitations, which on the gravity side correspond to quasi-normal modes (QNM)  of the $Z^2$ field on the D7-branes with very small imaginary parts compared to their real parts. The position of these peaks agrees almost exactly with the masses of mesons on a single D7-brane with the same value of $\Mq$ \cite{Kruczenski:2003be}. For $\epsilon=1$ no such peaks are present. This feature is straightforward to understand on the gravity side. As seen in \fig{embedding}, for small $\epsilon$ the branes' profiles approach those of \cite{Kruczenski:2003be}
everywhere except on a very thin throat that connects the branes  to the Poincar\'e horizon at the origin of AdS. The imaginary part of the QNMs measures the absorption probability by the horizon. Since this vanishes as the throat closes off, in this limit the imaginary parts of the QNMs frequencies vanish and the real parts converge to the real frequencies of \cite{Kruczenski:2003be}.  

\section{Discussion}\label{discussion}

We have considered $d=4$, $\mathcal{N}=4$, SU($\nc$) super Yang-Mills coupled to $\nf \ll \nc$  quarks, which breaks the supersymmetry to  $\mathcal{N}=2$. We have placed the theory at non-zero isospin and R-charge densities. We have shown that, if the conditions  \eqq{through1} and  \eqq{through2} are obeyed, then the ground state of the system is a supersymmetric, superfluid, color superconductor. The  solution 
that we considered is a unit-charge instanton centered at $r=0$. 
In this case the scale of gauge symmetry breaking is fixed by the charge densities. This is an essential difference with the physics of the Higgs branch\footnote{And in a certain approximation in the presence of a baryon density \cite{Ammon:2012mu}.} in which this scale is arbitrary.

Solutions for the case $\nf>2$ with any instanton number  are straightforward to construct because multidyonic instanton solutions for higher-rank gauge groups are known \cite{Eyras:2000dg}. For these configurations the possible symmetry-breaking patterns are richer  \cite{appear}. 

The superconducting property in our system is fundamentally different from that in what are usually referred to as ``holographic superconductors'', 
in which the broken symmetry is actually a global 
 symmetry \cite{Gubser:2008px,Hartnoll:2008vx,Hartnoll:2008kx}. In contrast,    in our case the broken color symmetry is a strongly coupled, non-Abelian gauge symmetry. In our probe approximation this breaking is encoded in the dissolution of some D3-brane charge on the D7-branes in the form of an instanton. The fact that the instanton size is completely fixed by the asymptotic charges proves that the Fermi seasickness \cite{Hartnoll:2009ns} that causes the dissolution of the D3-branes in the first place need not result in a runaway potential. Were we to include the backreaction of the D7-branes-plus-instanton  on spacetime  then the color breaking would be visible in the running of the color gauge group, as in \cite{Faedo:2016jbd}. In this backreacted scenario one would also be able to work in the grand-canonical ensemble for the R-charges. Instead, we are limited to the canonical ensemble because the gauge fields whose asymptotic values would allow us to define the R-charge chemical potentials are off-diagonal components of the ten-dimensional metric, whose dynamics is frozen in our probe approximation. 

To the best of our knowledge our solution is the first example of a supersymmetric color superconductor.  
The  BPS equations \eqq{BPS} are crucial to prove that the solution of the SYMH equations is also a solution of the full non-Abelian action of \cite{Myers:1999ps}. It would be interesting to investigate whether the full open string equations of motion are also satisfied, as in \cite{Thorlacius:1997zd}. 

Supersymmetry requires the isospin to be critical, \mbox{$\mui=\Mq$}, and the R-charges to be equal, $n_1=n_2$. It would be interesting to relax these conditions  both on the gravity side and on the field theory side. In the case of near-critical values it may be possible to compare a  weak-coupling field-theory analysis  along the lines  of \cite{Hollowood:2008gp,Hollowood:2011ep} with the strong-coupling holographic result.  

We have referred to our system as a superfluid because it breaks some global symmetries spontaneously. However, this does not directly imply that the system is able to support a superflow. It would be interesting to investigate this along the lines of 
e.g.~\cite{Amado:2013aea}.

In our solution the scalar and the time-component of the gauge field behave as 
\be
\phi=a_t=0+\mathcal{O}(r^2) 
\ee
in the IR.
The fact that the  scalar field vanishes implies that the induced metric on the D7-branes in the IR is $\mbox{AdS}_5 \times \mbox{S}^3$. The AdS$_5$ factor suggests that the IR physics may be conformally invariant despite the presence of several mass scales in the theory. The vanishing of the electric field suggests that relativistic invariance may also be restored in the IR despite the presence of non-zero charge densities. These speculations are under investigation \cite{appear}.

\section*{Acknowledgements}
We thank Stefano Carignano,  Roberto Emparan, Bartomeu Fiol, Eduardo Fraga, Jaume Garriga, Prem Kumar, and very specially Carlos Hoyos, for discussions. We are grateful to Jorge Casalderrey-Solana for a critical reading of the manuscript.  AF and DM are supported by grants FPA2016-76005-C2-1-P, FPA2016-76005-C2-2-P, 2014-SGR-104, 2014-SGR-1474, SGR-2017-754 and MDM-2014-0369. CP is supported by the STFC Consolidated Grant ST/P000371/1. JT is supported by the Advanced ARC project ``Holography, Gauge Theories and Quantum Gravity'' and by the Belgian Fonds National de la Recherche Scientifique FNRS (convention IISN 4.4503.15).




\end{document}